\documentclass[10pt,twocolumn,preprintnumbers,superscriptaddress,nofootinbib,aps,prd]
{revtex4-2}

\usepackage{graphicx}
\usepackage{amsmath}
\usepackage{srcltx}
\usepackage[utf8]{inputenc}
\usepackage[colorlinks]{hyperref}
\usepackage{url}
\usepackage{times}
\usepackage{slashed}
\usepackage{amssymb}
\usepackage{dsfont}
\usepackage{xcolor}
\usepackage{appendix}
\newcommand{\vev}[1]{\langle {#1} \rangle}
\newcommand{\lsim}{\lesssim}
\newcommand{\gsim}{\gtrsim}

\newcommand{\eq}[1]{Eq.~(\ref{#1})}

\newcommand{\ord}[1]{\mathcal{O}{(#1)}}
\newcommand{\beq}{\begin{equation}}
\newcommand{\eeq}{\end{equation}}
\newcommand{\bea}{\begin{eqnarray}}
\newcommand{\eea}{\end{eqnarray}}

\newcommand{\calG}{{\cal G}_0}

\newcommand{\mP}{M_{\rm P}}

\newcommand{\appropto}{\mathrel{\vcenter{
  \offinterlineskip\halign{\hfil$##$\cr
    \propto\cr\noalign{\kern2pt}\sim\cr\noalign{\kern-2pt}}}}}

\begin{document}

\pagestyle{plain}

\title{\boldmath Higgs Potential from Instantons}

\author{Hooman Davoudiasl}
\email{hooman@bnl.gov}

\author{Marvin Schnubel}
\email{mschnubel@bnl.gov}

\affiliation{High Energy Theory Group, Physics Department \\ Brookhaven National Laboratory,
Upton, NY 11973, USA}


\begin{abstract}

We propose that the Higgs potential, a key element in our understanding of Nature, is partially generated by the instantons of new confining dynamics, perhaps from a hidden sector. In this picture, while the Higgs itself is a fundamental field, it controls the strength of the non-perturbative interactions that give rise to its potential. This setup can be realized in ``Twin Higgs" hierarchy models, where the twin strong dynamics confines at or above the weak scale, but large quantum corrections to the Higgs potential are avoided. We examine a simple setup in which this instanton contribution is augmented by a quartic term, which is sufficient for a realistic electroweak symmetry breaking mechanism. The minimum of the potential is given by the Lambert $W_0$ function, with these assumptions. We discuss the predictions of this model and how it may be tested through measurements of the Higgs self coupling. Given the connection with non-trivial dynamics, one may also consider the prospects of accessing hidden sector states at colliders; this seems to be typically challenging in our setup. Symmetry restoration in the early Universe in this scenario is briefly examined. We also comment on the possible connection of our general setup with recent work on the physics of field space end points.

\end{abstract}
\maketitle


\section{Introduction}

The discovery of a scalar boson at the LHC \cite{ATLAS:2012yve,CMS:2012qbp} completed the menu of the particles contained in the Standard Model (SM). We will refer to this particle as the Higgs, however as the following  suggests its detailed properties may not coincide with that of the SM. While a number of key interactions for the Higgs have been measured to coincide closely with the SM predictions \cite{ParticleDataGroup:2024cfk}, its exact nature remains an open question. For example, the strength of the Higgs couplings to the first generation fermions are still unknown and pose a challenge for experiments. The self interactions of the Higgs have not been measured either and could potentially deviate from the SM expectations. Such a measurement would be important in establishing the form of the Higgs potential, which is key to answering fundamental questions about the mechanism for electroweak symmetry breaking (EWSB) \cite{Weinberg:1967tq}, the history of the Universe, and its likely destiny (see, {\it e.g.}, Ref.~\cite{Buttazzo:2013uya}).

\section{The Model}

The model we propose here assumes a new non-Abelian gauge interaction that we take to be an $SU(N)$ group, with $N\geq 2$.  This gauge sector may not have any direct connection with any of the SM interactions and could be part of a hidden sector. We do not specify the details of this sector, but it could possibly be where dark matter originates from. All we will assume for now is that there is a dimension-6 effective operator 
\beq
\delta {\cal L}_6=-\frac{H^\dagger H}{4 M^2} G_{\mu\nu}^a G^{a\mu\nu}\,,
\label{dim-6}
\eeq
where $H$ denotes the Higgs doublet field, $M\gg 100$~GeV is a high scale to be determined later, and $G_{\mu\nu}^a$ is the $SU(N)$ field strength tensor of adjoint index $a=1,2,\ldots,N^2-1$. The negative sign is required and presumably set by decoupled high energy physics.\footnote{Later, we will briefly discuss the implications of a positive sign in the context of a recent theoretical analysis in Ref.~\cite{Cheung:2024wme}.} Such operators could arise in a variety of SM extensions \cite{Curtin:2015fna}, such as the ``Twin Higgs" model \cite{Chacko:2005pe}. We will later comment in more detail on the differences in phenomenology of this model in our setup versus the original implementation. The normalization factor $1/4$ is chosen for later notational convenience.  We will adopt a convention in which the gauge kinetic term is of the form
\beq
-\frac{1}{4 g_0^2} G_{\mu\nu}^a G^{a\mu\nu}\,,
\label{g-kinetic}
\eeq
where the gauge coupling would be given by $g_0$ if the vacuum expectation value (vev) of the Higgs field $\vev{H} = 0$, {\it i.e.} when electroweak symmetry is unbroken. Taking into account the interaction in \eq{dim-6}, the effective gauge coupling $g$ at low energies is then given by 
\beq
\frac{1}{g^2} = \frac{1}{g_0^2} + \frac{\vev{H}^2}{M^2}\,,
\label{g2}
\eeq  
where consistency with observation requires $v\equiv \sqrt{2}\vev{H}\approx 246$~GeV.  

Based on general considerations, starting from some ultraviolet (UV) scale $\Lambda$, gauge interaction instantons can generate a scale $\mu$ given by 
\beq
\mu^4 \sim \Lambda^4 e^{-S}\,,
\label{mu4}
\eeq
were $S=8 \pi^2/g^2$ is the instanton action; see for example Refs.~\cite{Svrcek:2006yi,Hui:2016ltb} for discussions in the context of axion physics. Given the above dependence of $g$ on the value of the Higgs field, we then expect an instanton contribution to the Higgs potential $V(H)$. Let us assume that $V(H)$ also contains a quartic coupling of strength $\lambda$ and hence
\beq
V(H) = \mu_0^4 e^{-8\pi^2 \frac{H^\dagger H}{M^2}} + \lambda (H^\dagger H)^2\,.
\label{VH}
\eeq 
Here, $\mu_0$ is the scale generated by the initial value of the gauge coupling $g_0$.  In the appendix, we will briefly outline a scenario where the quartic term is replaced by the CP violating counterpart to the dimension-6 operator in \eq{dim-6}.

Let us define $x\equiv \vev{H}/M$. We note that the $SU(N)$ interactions become confining when $g^{-2} \ll 1$. With our assumptions, this requires both $g_0^{-2}\ll 1$ and $x^2\ll 1$. If these conditions are not satisfied, the scale of confinement for the hidden gauge dynamics would be pushed to well below the weak scale and a suitable Higgs potential would not be generated by the $SU(N)$ instantons. Here, we expect that the initial scale of confinement $\mu_0$ in \eq{VH} to be lowered to $\mu < \mu_0$ once the Higgs gets a vev and the electroweak symmetry is broken.  

In order to find the minimum of the Higgs potential, we solve $dV(x)/dx=0$. This yields 
\beq
y \, e^y = R\,,
\label{y-eq}
\eeq                              
where $y\equiv 8 \pi^2 x^2$ and 
\beq
R\equiv \frac{2}{\lambda}\left(\frac{2 \pi \mu_0}{M}\right)^4\,.
\label{R}
\eeq
The solution to \eq{y-eq} is given by Lambert function $W_0$ \cite{lambert1758observationes}, with 
\beq
y = W_0(R)\,,
\label{W0}
\eeq
for $R\geq 0$, which holds for all cases of interest in our work.  

Here, we would like to note  that this simple treatment of electroweak symmetry breaking in our scenario ignores the feedback of a non-zero Higgs vev on the potential in \eq{VH}. That is, with our assumed interactions, we expect that after the Higgs field condenses, the effective $SU(N)$ coupling gets smaller. This in turn suppresses the instantons, leading to a smaller  scale $\mu$ in the Higgs potential, which leads to a smaller value of $\vev{H}$. A smaller vev for the Higgs makes the gauge coupling larger and enhances the instanton effect, and the process is repeated. This iterative picture is in reality a smooth process where $\vev{H}$ evolves from a larger initial value to a final one that is somewhat smaller. The process is converging, and we find for the final vev $v=\sqrt{2}M x_f$ and instanton generated scale $\mu_f$
\begin{eqnarray}
    (4\pi x_f)^2&=W_0\left(2R\right) \label{realvev}\\
    \mu_f^4&=\mu_0^4e^{-8\pi^2 x_f^2}\,. \label{finalmu}
\end{eqnarray}
Note that there is also the effect of the decrease in the cosmic temperature which moves the vev in the opposite direction, making the instantons stronger and $\vev{H}$ larger.  However, if the potential in \eq{VH} is taken to correspond to that obtained well after $SU(N)$ confinement, we may assume that the instanton strength is effectively frozen and not affected by further thermal evolution of the gauge interactions.

\section{Phenomenology}

\begin{figure}[t]\vskip0.25cm
	\centering
	\includegraphics[width=\columnwidth]{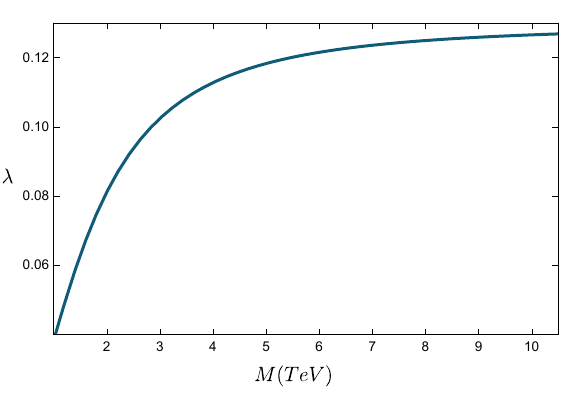}
	\caption{Value of the Higgs self coupling parameter $\lambda$ which yields the experimentally determined Higgs vev and mass, as a function of the mass $M$ where the new physics is expected to set in.}
	\label{lam-M}
\end{figure}

Using \eq{realvev}, we may now solve for $\mu_0$ in terms of $\lambda$ and $M$, by setting the Higgs vev $v$ to its experimentally determined value. We will denote the observed Higgs boson by $h$, and $H\to (v + h)/\sqrt{2}$ after EWSB. Substituting this expansion into $V(H)$, the Higgs boson mass $m_h$ is then defined by 
\beq
m_h^2 \equiv \frac{d^2 V(h)}{dh^2}|_{h=0}\,,
\label{mh2}
\eeq
which we set to its measured value $m_h = 125.2$~GeV \cite{ParticleDataGroup:2024cfk}. We can also obtain $\mu_0$ in terms of $\lambda$ and $M$ from the above equation and \eqref{finalmu}.  With the above solution for $\mu_0$, we may then obtain values of $\lambda$ which yield the experimentally determined Higgs parameters, as a function of $M$; this is plotted in Fig.~\ref{lam-M}.

\begin{figure}[t]\vskip0.25cm
	\centering
	\includegraphics[width=\columnwidth]{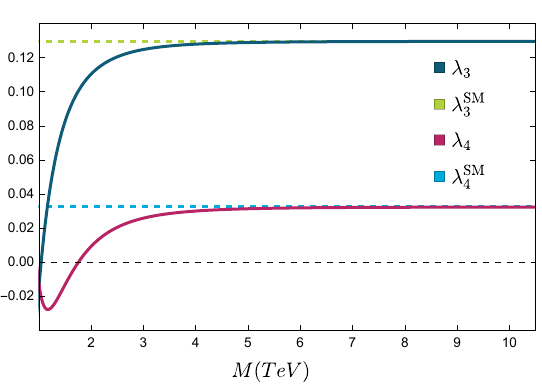}
	\caption{The cubic and quartic couplings of the Higgs boson in the scenario presented in this work are represented by the solid magenta and dark blue curves, respectively.  The dashed green and light blue lines represent, respectively, the corresponding SM self coupling predictions of the Higgs boson. As the mass scales gets higher, the parameters in our model approach their SM values.}
	\label{lam34}
\end{figure}

Given that the Higgs potential introduced here is different than that of the SM, it would be interesting to distinguish the two cases.  Here, we focus on the value of cubic and quartic couplings of the Higgs boson $h$, denoted by $\lambda_3$ and $\lambda_4$.  Note that the cubic coupling has dimension 1 and hence we define $\lambda_3$ as the coefficient of $v h^3$ in the broken phase, while $\lambda_4$ is the coefficient of the $h^4$ term.  These coefficients are then given by
\beq
\lambda_3 = \frac{1}{3!\, v} \frac{d^3 V(h)}{dh^3}|_{h=0} \quad \text{and} \quad 
\lambda_4 = \frac{1}{4!} \frac{d^4 V(h)}{dh^4}|_{h=0}.
\label{lam3}
\eeq
We have plotted $\lambda_3$ and $\lambda_4$ in Fig.~\ref{lam34}. 

As we can see the cubic coupling of the Higgs boson approaches that of the SM prediction as $M$ gets larger.  For $M\lsim 3$~TeV, a measurement of this quantity at the $\sim 10\%$ level can uncover a deviation from the SM\footnote{A similar result is found in Ref. \cite{Chun:2019box}, which however takes $M\sim300$~GeV and therefore it is not clear to us whether the EFT assumption still holds for multi Higgs final states requiring energies above $M$. Additionally, we want to emphasize that expanding the exponential in the effective Higgs potential in general does not capture the full extent of the physics at play.}. This may be feasible at a future collider through measurements of double Higgs production \cite{Narain:2022qud}. Similarly, the quartic coupling of the Higgs boson agrees with its SM value in the limit of large $M$, as expected. Since it is extracted from triple Higgs cross sections, its measurement is experimentally challenging.  Nonetheless, given the large deviation from the SM value for smaller scales $M\lesssim3$~TeV, a precision measurement is not required and a detection of the quartic coupling anywhere near the SM expectation can potentially rule out this model.

\begin{figure}[t]\vskip0.25cm
	\centering
	\includegraphics[width=\columnwidth]{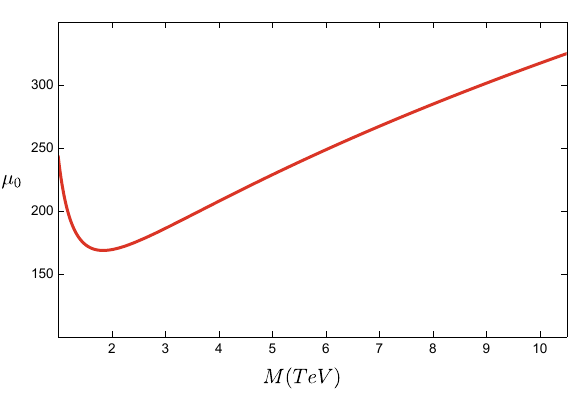}
	\caption{The Higgs potential parameter $\mu_0$ vs. $M$.}
	\label{mu0-plot}
\end{figure}

Let us now address the question of Higgs potential stability in our model.  We note that the first term in \eq{VH} for $V(H)$ is assumed to be generated by non-perturbative effects.  The scale of this dynamics is $\sim \mu_0$, and as can be seen from Fig.~\ref{mu0-plot} $\mu_0\sim \ord{v}$.  As one probes higher energy scales, this contribution is expected to become small, as the instantons are suppressed with decreasing $g_0$.  

\begin{figure}[t]\vskip0.25cm
	\centering
	\includegraphics[width=\columnwidth]{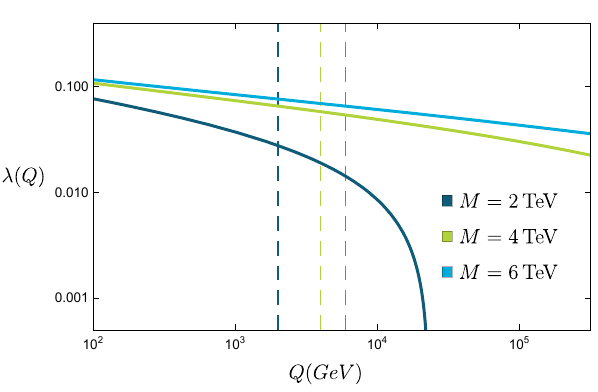}
	\caption{Running of $\lambda$ for three values of $M$.  The dark blue, green and light blue curves correspond to $M=2,4,6$~TeV, respectively. The vertical, dashed lines represent the mass scale with the same color coding. Above this respective scale the explicit UV completion needs be taken into account because the EFT assumption is no longer valid.}
	\label{lambdaRun}
\end{figure}

Hence, for scales $Q \gg v$, relevant to the question of stability, we may ignore the instanton contribution and focus on the second term proportional to $\lambda$, whose value as a function of $M$ is plotted in Fig.~\ref{lam-M}.  We can calculate the running of $\lambda$ with $Q$, based on the well known renormalization equations, which apart for the initial value of $\lambda$ at $Q_0=v$ are the same as in the SM, which is currently known up to three-loop order \cite{Chetyrkin:2013wya}.  We have 
\bea\nonumber
16 \pi^2 \frac{d \lambda}{d t} &=& 24 \lambda^2 + 
\lambda (12 y_t^2 - 9 g_2^2 - 3 g_1^2) - 6 y_t^4\\ 
&+& \frac{9}{8} g_2^4 +\frac{3}{8} g_1^4 + \frac{3}{4} g_2^2 g_1^2\,,
\label{lambda-running}
\eea
where $y_t$ is the SM top Yukawa coupling; $g_2$ and $g_1$ are the $SU(2)_L$ and $U(1)_Y$ gauge couplings, respectively.  We have plotted the running of $\lambda$ with $Q$, for a few values of $M$, in Fig.~\ref{lambdaRun}. The coupling stays positive and does not destabilize the Higgs potential up to energy scales $Q\sim M$. For higher energy scales, the EFT assumption is no longer valid and the theory must be matched onto a concrete UV completion which will also change the running of $\lambda$ for $Q>M$.      

\section{Cosmological Considerations}

So far, we have been discussing the parameters required for realistic electroweak symmetry breaking, and key  features of the Higgs potential in our model.  However, here we would like to discuss the general aspects of this theory in the early Universe, in particular at temperatures above the weak scale $T\gsim v$.  

We start by asking whether the hidden gauge dynamics can be expected to be in thermal equilibrium with the SM at such temperatures.  The most direct link between these two sectors is through the assumed dimension-6 operator in \eq{dim-6}.   We may ask at what temperature the hidden $SU(N)$ gauge theory decouples from the SM.  This is governed by the Hubble rate ${\cal H} \sim g_*^{1/2} T^2/\mP$, where $g_*\sim 100$ is the relativistic degrees of freedom and $\mP\approx 1.2 \times 10^{19}$~GeV is the Planck mass.  Based on dimensional analysis, the decoupling temperature $T_d$ is given by
\beq
\frac{T_d^5}{M^4} \lsim {\cal H}(T_d)\,.
\label{Td}
\eeq     
We have largely assumed $M\leq 10$~TeV, and hence we find the upper bound on $T_d$ for our parameter space of interest is typically $T_d\lsim 0.2$~GeV.  Hence, for $T$ at or above the weak scale we can expect the hidden gluons to be in thermal equilibrium with the SM.  

Given the above analysis, we expect that for $T\gsim \mu_0$, the hidden $SU(N)$ theory enters the weak coupling regime, due to asymptotic freedom.  This suggests that the first term in $V(H)$, contributed by instantons, becomes suppressed and for $T\gg \mu_0$, {\it i.e.} for temperatures well above the weak scale, the Higgs potential is dominated by the quartic term, which has a minimum at $\vev{H}=0$.    Thermal loops, which yield a contribution to the potential $\propto T^2 H^\dagger H$ only enhance this effect and therefore at $T\gg \mu_0$ we may expect that the electroweak symmetry is restored, which also coincides with the deconfined phase of the hidden $SU(N)$ theory.  

In the limit that the hidden gluons are the only states from that sector which are in thermal equilibrium, based on general arguments, we may expect that as the $SU(N)$ theory confines at $T\lsim \mu_0$, it would go through a first order phase transition \cite{Svetitsky:1982gs,Yaffe:1982qf}.  This transition sets up the Higgs potential through non-perturbative instanton interactions.  Therefore, EWSB in our theory is expected to be immediately preceded by a first order phase transition at $T\sim \mu_0$, which could give rise to observable primordial gravitational waves \cite{Schwaller:2015tja,Caprini:2019egz}.

In the preceding discussion, we have implicitly assumed that the hidden gluons could have been produced at high temperatures prior to getting confined.  The confinement would lead to the appearance of hidden ``glue-balls" that are analogues of those from quantum chromodynamics (QCD) in the SM.  Given their strong interactions, we may expect that the hidden glue-balls would annihilate and drive their population to the lightest such state, which is assumed to be a $0^{++}$ scalar state of even parity and charge conjugation \cite{Curtin:2015fna} that can mix with the Higgs and decay through it; we will denote this state by ${\cal G}_0$.  In the parameter space typical of our model, we expect the mass of $\calG$ to be large compared to the Higgs mass: $m_{\calG} \gg m_h$.  Hence, its contributions to the  Higgs exotic decays would be negligible, only allowed as an off-shell state.  However, the decay of $\calG$ through the Higgs into the SM sector would typically be prompt.  This ensures that the hidden glue-balls would not form a late decaying relic population, which could disrupt Big Bang Nucleosynthesis.  Hence, we do not expect any constraints from cosmology on our scenario. 

\section{Hidden Sector Signals}

Given that the lightest hidden glue-ball $\calG$ can mix with the Higgs, we will next consider the prospect for discovering it in future collider measurements.  We may adapt the results in Ref.~\cite{Curtin:2015fna} to get an estimate for the typical size of the mixing angle between the Higgs boson $h$ and $\calG$.  To do so, we will assume that hidden dynamics is governed by $SU(3)_X$, 
in order to facilitate analogy with insights from the SM strong interactions of low energy QCD.  

Let us rewrite the interaction in \eq{dim-6}, by making the dependence on the hidden fine structure constant $\alpha_X$  explicit              
\beq
\delta{\cal L}_6 \to -\alpha_X \frac{H^\dagger H}{2\,M_X^2} G_{\mu\nu}^a G^{a\mu\nu}\,,
\label{dim-6-v2}
\eeq
where $M_X$ is the scale of physics well above the confinement scale of the hidden sector, which we will take to be $\mu_0$.  In our work, we typically have $\mu_0 \gsim \ord{\rm 100~GeV}$ and $M_X\gsim \ord{\rm TeV}$, which furnishes the assumed hierarchy of scales.  Let us denote the decay matrix element for $\calG \to \text{`vacuum'}$ by $F_0$, where     
\beq
F_0\equiv \frac{1}{2}\langle 0| G_{\mu\nu}^a G^{a\mu\nu} |0^{++}\rangle\,.
\label{F0def}
\eeq
We may take \cite{Curtin:2015fna} 
\beq
4 \pi \, \alpha_X F_0 \approx 2\, m_{\calG}^3\,.
\label{F0val}
\eeq 
With the above definitions, \eq{dim-6-v2} would then lead to the mass mixing term
\beq
\delta{\cal L}_6 \to - \Delta^2 \hat{h}\, \hat{\calG} + \ldots\,,
\label{mass-mixing}
\eeq
where hatted fields are in interaction basis and we have 
\beq
\Delta^2 \equiv \frac{(4 \,\pi\, \alpha_X F_0)v}{4\, \pi M_X^2}\,.
\label{Delta2}
\eeq 

One can make a transformation to the mass basis, with the Higgs boson $h$ and hidden glue-ball $\calG$ as the propagating states.  We have
\beq
\hat{h} = \cos \phi\, h + \sin \phi\, \calG
\label{hhat}
\eeq
and
\beq
\hat{\calG} = - \sin \phi\, h + \cos \phi\, \calG \,,
\label{G0hat}
\eeq
where the mixing angle $\phi$ is given by 
\beq
\tan 2\phi = \frac{2 \Delta^2}{m_{\calG}^2 - m_h^2}.
\label{tan2phi}
\eeq
In the above, $m_h=125.2 \pm 0.11$~GeV is the Higgs boson mass \cite{ParticleDataGroup:2024cfk}.  For $SU(3)_X$, we adopt $m_{\calG}\sim 7 \mu_0$ \cite{Curtin:2015fna}, which typically implies $m_{\calG}\gsim 1$~TeV.  

Assuming $m_{\calG}\sim M_X$, we then roughly estimate that $\phi\lsim 4\times 10^{-2}$.  This implies a shift in the Higgs boson couplings at the level of $\ord{\phi^2/2}\lsim 8 \times 10^{-4}$, which is well below the current precision of few \% or worse.  It is not clear that this level of precision in the couplings of $h$ would be achievable in the foreseeable future \cite{Narain:2022qud,Forslund:2023reu}.  One may consider the production of $\calG$ on resonance through gluon fusion.  This can be estimated from the production of a Higgs boson with mass $m_h \to m_{\calG}$, in gluon fusion, but suppressed by $\phi^2$.  Again, the level of expected suppression makes this a challenging prospect. We would next like to briefly mention potential UV models that would generate the effective operator \eqref{dim-6}.

Here, we also add that the interaction we have assumed in \eq{dim-6}, could in principle lead to the emergence of a dimension-6 operator in the SM effective field theory of the form $ C_{H\square}H^\dagger H \square H^\dagger H/\Lambda^2$, suppressed by the scale $\Lambda$ which we take to be 1 TeV \cite{Ahmed:2024hpg}.  Following the treatment in Ref.~\cite{Ahmed:2024hpg}, one can show that in our scenario, where the lightest composite state has a mass $m_{\calG}\gsim 1$~TeV, the Wilson coefficient $C_{H\square}\lsim \text{few}\times 10^{-2}$, which does not lead to a significant constraint on our model, given the current experimental bounds \cite{Ahmed:2024hpg}.

\subsection{A More Hidden Twin Higgs Scenario} 

At first, we focus on the previously mentioned Twin Higgs model designed to solve the hierarchy problem up to some cutoff scale $\Lambda>\Lambda_{\text{EW}}$ known as the ``little hierarchy problem". In the minimal version, this is done through realizing the experimentally observed Higgs boson as the Goldstone boson of an approximate, global $SU(4)$ symmetry. The UV sensitivity of the Higgs mass is ameliorated by extending the fermion content with partner states which are related to the SM fermions by a $\mathds{Z}_2$ symmetry \cite{Craig:2015pha}. Here, we would like to add an additional twist to the original idea. Assuming that the Twin $SU(3)_D$ coupling constant is given by $g_D=1.5\times g_3$ at a scale $\mu=1$~TeV and no dark colored states are lighter, we find that the dark QCD develops an IR Landau pole and confines at $\mu_0\sim250$~GeV, which is in accordance with our assumptions; see Fig. \ref{mu0-plot}. The little hierarchy problem is then solved up to scales $\Lambda\sim5$~TeV allowing for a $\mathcal{O}(15\%)$ fine-tuning. We note that while the Twin  scenario can lead to the cancellation of  contributions from  the SM and UV physics to the Higgs mass parameter, which by the way also could justify assuming a small mass tern, which we have ignored.  Such a  suppression does not apply to the strength of the dimension-6 operator in \eq{dim-6}, which is  mediated only by the hidden sector states.

A major phenomenological difference is that with the above mentioned changes there are no light new states. The Higgs boson couplings due to mixing with the lightest dark glueball are only slightly altered. Furthermore this glueball is too heavy to be produced on-shell in Higgs decays, and so there are no expected events with missing final state energy at the LHC. Hence, the only deviations from a SM-like Higgs lie in the self-couplings. To summarize, all essential properties of the Twin Higgs model remain preserved if one takes the dark QCD to confine in the way presented above, while simultaneously yielding a Higgs boson that looks like the one experimentally observed, thus providing additional viable parameter space for this model.

\subsection{Other Realizations}

\begin{figure}[t]\vskip0.25cm
	\centering
	\includegraphics[width=.6\columnwidth]{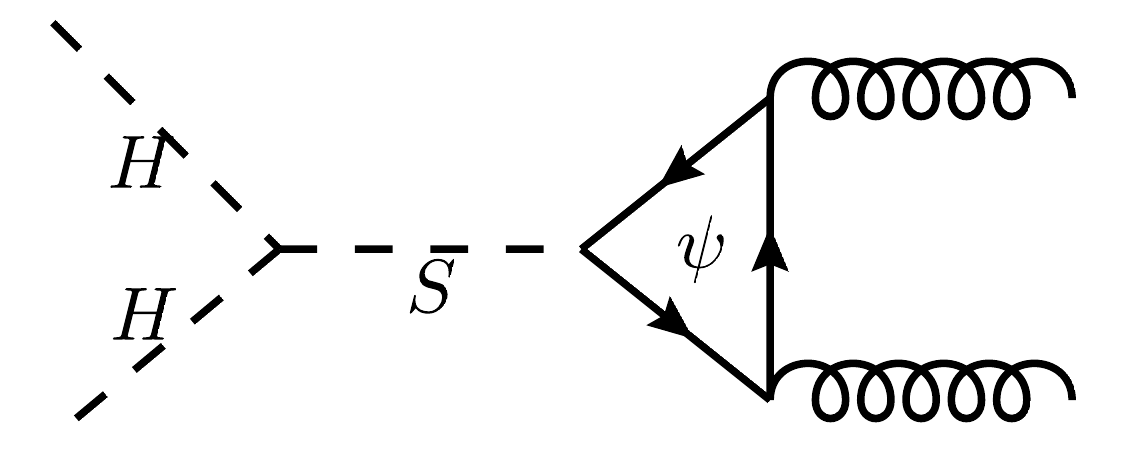}
	\caption{Feynman diagram generating the effective operator \eqref{dim-6} after integrating out the heavy dark scalar $S$ and fermion $\psi$ running in the loop.}
	\label{feynmandiag}
\end{figure}

In addition to the Twin Higgs framework, here we present some alternative UV realizations, as well as comment on their impact on the question of stability of the Higgs potential. At tree-level, there is no UV completion that could generate the required dimension-6 operator. At the one-loop level, we find that a dark scalar $S$ which couples to dark fermions $\psi$ that carry $SU(N)$ color can generate the effective interaction \eqref{dim-6} through the Lagrangian \cite{deBlas:2017xtg}
\beq
\mathcal{L}\supset
- y_\psi S \bar \psi \psi - \lambda_S S^2(H^\dagger H)+\bar{\psi}(i\slashed{D} -  m_\psi)\psi\,.
\label{Sint}
\eeq
The Feynman diagram responsible for generating the effective operator is shown in Fig.~\ref{feynmandiag}. The effective mass scale $M$ in \eqref{dim-6} is then given by
\beq
\frac{1}{M^2}=\frac{\alpha_X}{3\pi}\frac{\lambda_S\, y_\psi\, v_S}{M_S^2 \,(m_\psi + y_\psi\, v_S)}\,,
\label{M2}
\eeq
with $m_{S,\psi}$ the UV mass of the dark scalar and the fermion, respectively, $v_S$ the vev of $S$, and $y_\psi$ the Yukawa coupling of $\psi$ to $S$. Additionally,  this new scalar particle could potentially stabilize the Higgs potential in the absence of further new particles that couple to the Higgs. This is fulfilled for a mass of $M_S\sim1$~TeV and scalar-Higgs couplings $9\times10^{-4}\leq \lambda_S/(4\pi)^2\leq2\times10^{-2}$ \cite{Hiller:2024zjp}.  Experimental bounds on the allowed mixing of the Higgs boson with $S$ suggest $\lambda_S\lsim 0.2$ \cite{Hiller:2024zjp}, for $M_S\sim 1$~TeV.

Other single particle extensions of the SM that generate \eqref{dim-6} at the one-loop level under the assumption that the coupling to QCD and the dark QCD are of comparable size and structure include the leptoquarks $\omega_{1,2,4}$, $\Pi_{1,7}$ and $\zeta$ \cite{Dorsner:2016wpm}, the diquarks $\Omega_{1,2,3}$ and $\Upsilon$ \cite{Giudice:2011ak,Englert:2024nlj,Marciano:1980zf}, the Manhohar-Wise model \cite{Manohar:2006ga}, the singlet vector-like quarks (VLQ) $U$, $D$ and $Q_{1,5,7}$ \cite{Aguilar-Saavedra:2009xmz}, and the triplet VLQs $T_{1,2}$ \cite{Gargalionis:2024jaw}. The corresponding effective mass scale can be extracted from the respective SM effective field theory (SMEFT) Wilson coefficients found in Refs. \cite{Gargalionis:2024jaw,Guedes:2023azv,Guedes:2024vuf}. Out of these, the models $\omega_1$, $\Pi_1$, $\zeta$, $D$, $U$, $Q_1$ and $Q_5$ also stabilize the Higgs potential at the Planck scale \cite{Bandyopadhyay:2016oif,Bandyopadhyay:2021kue,Hiller:2022rla,Hiller:2024zjp}. Note that models $D$ and $U$ require particle multiplicities $N_F\geq2$, while $Q_5$ demands $N_F=1$.  We refer the interested reader to the above cited works for more details.

Before concluding this work, we will make a connection to recent work in Ref.~\cite{Cheung:2024wme} on the physics of end points in field space.  Note that the sign of the operator in \eq{dim-6} is determined by the UV physics generating this operator, which depending on the specific UV realization can be either positive or negative. This, for example, can be easily seen using our simple model leading to \eq{M2}, where the sign of the product $\lambda_S \, y_\psi$ is not fixed and could result in either sign for the Wilson coefficient of the dimension-6 interaction.

Here, we note that if the interaction in \eq{dim-6} has a positive sign, it would lead to an instanton produced potential of type 
\beq
V(H) \propto e^{
	\frac{|H|^2 - \overline{|H|^2}}{\Lambda^2}
}\,,
\label{VH-II}
\eeq
with $|H|^2 \equiv H^\dagger H$ and $\Lambda^2\equiv M^2/(8 \pi^2)$.  The end point of the field configuration is given by 
\beq
\overline{|H|^2} = \frac{8\pi^2\Lambda^2}{g_0^2}. 
\label{endpoint}
\eeq
The discussion in Ref.~\cite{Cheung:2024wme} identifies the true cutoff of the theory underlying this potential to be of order $\Lambda$, which can only be manifested by a coherent field configuration with a large number of bosons.  

In the context of our model, the above endpoint for $H$ is related to the size of the hidden gauge theory coupling $g_0$.  Obviously, if $|H|^2$ gets too large, the kinetic term for the hidden gauge field would have the wrong sign and the theory becomes ill-defined.  In this framework, the scalar field endpoint then corresponds to the regime of validity of the gauge theory that couples to the Higgs bi-linear.  For a weakly coupled gauge sector, the endpoint $\overline{|H|}$ in \eq{endpoint} can be large.  However, the true cutoff $\sim \Lambda$ would be much lower, but not easily accessible to low energy experiments.  

\section{Summary and concluding remarks}  

In this work, we considered the possibility that the Higgs potential is generated in part by non-trivial instanton dynamics from a new sector, leading to electroweak symmetry breaking.  This sector may be unrelated to gauge sector of the SM.  In this scenario, the Higgs is a fundamental field, however its background value contributes to the strength of the new confining gauge theory.  We found that a feasible path to testing this scenario is through precision measurements of the Higgs boson self coupling at future colliders.  Access to the composite states of the new gauge sector seems unlikely in the basic model we have proposed here.  The confinement phase transition in the new sector is expected to be first order, which can lead to potentially detectable gravitational waves at future observatories.  In general, there are no severe cosmological constraints on our proposed setup.  We also made connections with the physics of field space endpoints, in the context of the general effective interactions we have introduced.

\begin{acknowledgments}
We thank R. Szafron for discussions on related topics.  This work is supported by the US Department of Energy under Grant Contract DE-SC0012704. M.S. gratefully acknowledges support from the Alexander von Humboldt Foundation as a Feodor Lynen Fellow.
\end{acknowledgments}

\appendix
\section{Alternative CP Violating Setup}

Here, we will examine  a scenario where the Higgs potential is entirely  generated by CP conserving and CP violating dimension-6 operators.  Thus, in addition to the interaction in \eq{dim-6}, we will also consider the following operator 
\beq
\frac{H^\dagger H}{M'^2} G_{\mu\nu}^a \tilde{G}^{a\mu\nu}\,,
\label{dim-6-CPV}
\eeq
where $\tilde{G}^{a\mu\nu}$ denotes the dual field strength tensor, and $M'$ is a new effective field theory (EFT) scale.  The above interaction is CP violating and can be generated if there are complex phases in the UV theory.  Once $\vev{H}\neq 0$, \eq{dim-6-CPV} contributes to an effective ``$\theta$ angle" in the low energy theory.  On general grounds, the Higgs potential generated by the dimension-6 operators is of the form 
\beq
V(H) = \mu_0^4 \,e^{-8\pi^2 \frac{H^\dagger H}{M^2}}
\cos\left(\theta +\frac{H^\dagger H}{M'^2}\right)\,,
\label{VH-CPV}
\eeq
where $\theta$ is the CP violating parameter of the hidden $SU(N)$ gauge theory.  

One can straightforwardly show that the extrema of the above potential are obtained from the equation 
\beq
\tan\left(\theta +\frac{v^2}{2 M'^2}\right) = - \frac{8 \pi^2 M'^2}{M^2}.
\label{minima}
\eeq
In the following, we will consider a few special cases, assuming a positive coefficient for the operator (\ref{dim-6-CPV}).

\begin{itemize}
\item {\it (i)} $\theta = 0 $ and $M=M'$: In this simplest case, \eq{minima} yields $v^2 \approx \pi M^2$.  Since the validity of the EFT implies $v\ll M,M'$, we do not consider this case as a viable option in our scenario.  We will thus assume that  the EFT is a good representation of the new physics and $v\ll M'$, which suggests that the contribution of $v^2/(2 M'^2)$ to the tangent in \eq{minima} is small.   

\item {\it (ii)} $\theta\neq0$: To have a valid EFT description, we need to require $v^2/(2M'^2)\ll1$. Since the right-hand side of equation \eqref{minima} is always negative, this means that $\theta\approx\pi/2$, independent of the two mass scales. For every sufficiently high mass scale $M'$, we can then determine the amount that $\theta$ can deviate from $\pi/2$ such that $\theta+v^2/(2M'^2)\geqslant\pi/2$. Solving Eq. \eqref{minima} then gives the corresponding mass scale $M$. Note that requiring that the vev sits at the minimum of the potential excludes other solutions. In the simplest realization of this case we choose $\theta=\pi/2$ and $M'=M$. The model then has only two free parameters, $M$ and $\mu_0$, and we can determine their values by requiring that the vev and the Higgs mass agree with their experimentally observed values. Hence we find $M\approx1547$~GeV and $\mu_0\approx475$~GeV. With these values the predictions for the cubic and quartic Higgs self-couplings are $\lambda_3\approx-0.043$ and $\lambda_4\approx-0.097$, respectively, which is still within the current experimental uncertainty.


\end{itemize}

We hence conclude that in the scenario considered in this appendix, the hidden  instantons can  generate a potential for the Higgs if the UV model possesses unsuppressed CP violation manifested in a maximal $\theta$-angle.  


\bibliography{H-Inst-refs.bib}

\end{document}